\documentclass[fleqn,usenatbib]{mnras}

\usepackage{newtxtext,newtxmath}

\usepackage[T1]{fontenc}
\usepackage{ae,aecompl}
\usepackage{booktabs}

\usepackage{graphicx,subfig}	
\usepackage{amsmath}	
\usepackage{amssymb}	
\usepackage{bm}
\usepackage{float}

\renewcommand{\(}{\left(}
\renewcommand{\)}{\right)}
\renewcommand{\[}{\left[}
\renewcommand{\]}{\right]}

\newcommand{\ba}{\begin{eqnarray}}
\newcommand{\ea}{\end{eqnarray}}
\newcommand{\labeq}[1] {\label{eq:#1}}
\newcommand{\refeq}[1] {(\ref{eq:#1})}
\newcommand{\labfig}[1] {\label{fig:#1}}
\newcommand{\reffig}[1] {\ref{fig:#1}}
\newcommand{\labsec}[1] {\label{sec:#1}}
\newcommand{\refsec}[1] {\ref{sec:#1}}

\newcommand{\hatC}{\ensuremath{\hat{C}}}

\newcommand{\mbC}{\ensuremath{\mathbf{C}}}
\newcommand{\mbhatC}{\ensuremath{\mathbf{\hatC}}}
\newcommand{\tr}{\ensuremath{^T}}

\title[Understanding parameter differences]{Understanding parameter differences
between analyses employing nested data subsets}

\author[S.\ Gratton and A.\ Challinor]{
Steven Gratton$^{1,2}$\thanks{Contact e-mail: \href{mailto:stg20@cam.ac.uk}{stg20@cam.ac.uk}} and 
Anthony Challinor$^{1,2,3}$\thanks{Contact e-mail: \href{mailto:a.d.challinor@ast.cam.ac.uk}{a.d.challinor@ast.cam.ac.uk}} \\
$^{1}$Institute of Astronomy, Madingley Road, Cambridge CB3 0HA, UK\\
$^{2}$Kavli Institute for Cosmology Cambridge, Madingley Road, Cambridge CB3 0HA, UK\\
$^{3}$DAMTP, Centre for Mathematical Sciences, Wilberforce Road, Cambridge CB3 0WA, UK
}
\date{18 November 2019}

\pubyear{2019}

\begin{document}
\label{firstpage}
\pagerange{\pageref{firstpage}--\pageref{lastpage}}
\maketitle

\begin{abstract}
We provide an
analytical argument for understanding the 
likely nature of parameter shifts between those 
coming from an analysis of a dataset and from a subset of that dataset, assuming
differences are down to noise and any intrinsic variance alone.  This
gives us a measure against which we can interpret changes seen 
in parameters and make judgements about the coherency of the data 
and the suitability of a model in describing those data.   
\end{abstract}


\section{Introduction}

One would typically expect the posterior distributions of
the parameters of a model to change
as the datasets used to constrain them are
changed.   A part of this must be attributable to ``scatter'', i.e., noise
and any intrinsic variance assumed by the model.
However, it could also  be an indication of a problem, 
either in the data (e.g., a systematic error in one of the
datasets or an unacounted-for relative calibration 
between two datasets) or in the model (i.e., the model 
is incomplete and unable to well-describe all of the data).   

In this paper we provide an
analytical argument for understanding the 
likely nature of parameter differences in the ideal, scatter-only,  case.   This
gives us a measure against which we can interpret changes seen 
in parameters inferred from using subsets of the data.  Thus we can make 
judgements about the internal coherency of the data 
and the appropriateness of a model for describing those data.

In an appendix we show how the approach presented here can be used
quickly to rederive the result of \citet{wilks} involving the relation between the goodness-of-fits 
of a standard analysis and one with additional model parameters. 

Some of the techniques described in detail in this paper were used in 
\citet{2016A&A...594A..11P,2017A&A...607A..95P,2019arXiv190712875P}.  For similarly-motivated
work see \citet{2019PhRvD..99d3506R,2019arXiv191007820L}.

\section{Derivation of Main Result}
\labsec{thearg}

Imagine one has a collection of data, denoted by the vector $\hat{X}$.  One has
a parametric model in mind to describe these data, depending on a 
collection of parameters denoted by a vector $P$.  Let us write the probability density 
for a realisation $X$ of the data under the model as
\begin{equation}
p(X) dX=e^{-S} dX, \label{eq:actdef}
\end{equation}
where $S$ is a function of $X$ and the parameters $P$ and $dX$ is the appropriate multi-dimensional
measure on the data.  

Let us now assume that the model can indeed describe the data and
  that the true parameter values are $P_0$.
Let us expand $S$ to
second order in the parameters about $P_0$:
\begin{align}
S&=S_0+S'^T \delta P+ \frac{1}{2} \delta P^T S'' \delta P+\cdots
   \nonumber \\
&\approx  S_0+S'^T \delta P+ \frac{1}{2} \delta P^T \overline{S''}
  \delta P \nonumber \\
&= \frac{1}{2} \left(\delta P+\overline{S''}^{-1} S' \right)^T
  \overline{S''} \left(\delta P+\overline{S''}^{-1} S' \right) 
+\, \mathrm{const}.
\label{eq:actexp}
\end{align}
Here $S'$ denotes the vector of derivatives of $S$ with respect to the parameters, and $S''$
denotes the matrix of second derivatives.
In
the second line we have made the (typically-good) approximation of replacing $S''$ with its average 
$\overline{S''}$, where here and onwards an overline denotes an ensemble average of the
indicated object with respect to data realizations from the assumed model 
with parameters $P_0$.  Equation (\ref{eq:actexp}) motivates the ``maximum-likelihood'' estimator
for the parameters, and we now briefly recap some of its properties in preparation for
what will follow.  The fluctuation in parameters around $P_0$ for any
given realisation of the data is given by
\begin{equation}
\delta P=-\overline{S''}^{-1} S', \label{eq:dpee}
\end{equation}
where $S'$ is evaluated for the realization in question.  This can be seen to be unbiased as follows.
From Eq.\ (\ref{eq:dpee}) we have $\overline{\delta P} =
-\overline{S''}^{-1} \overline{S'} $. Using  Eq.\ (\ref{eq:actdef})
and considering, for example, the $i$th parameter $P^i$, we have
\begin{align}
\overline{S_{,i}} &= \int dX S_{,i} e^{-S} \nonumber \\
&= -\int dX \left(e^{-S} \right)_{,i} \nonumber \\
&= -\left(\int dX e^{-S} \right)_{,i} \nonumber \\
&= -1_{,i}=0,
\end{align}
where $S_{,i} = \partial S / \partial P^i$. 
In the last line we have used the fact that the probability
distribution for $X$ is normalized to unity.  It follows that
$\overline{\delta P} = 0$.

To obtain the covariance of the
parameters we average the outer product of Eq.\ (\ref{eq:dpee}) over the ensemble.  For this
we need
\begin{align}
\overline{S_{,i} S_{,j}} &= \int dX S_{,i} S_{,j} e^{-S} \nonumber \\
&= -\int dX S_{,i}\left(e^{-S} \right)_{,j} \nonumber \\
&= -\left(\int dXS_{,i}e^{-S} \right)_{,j}+\int dX S_{,ij} e^{-S}
  \nonumber \\
&= 1_{,ij}+\overline{S_{,ij}} \nonumber \\
&= \overline{S_{,ij}},
\end{align}
and hence
\begin{equation}
\overline{\delta P \delta P^T}=\overline{S''}^{-1} \overline{S''} \, \overline{S''}^{-1}= \overline{S''}^{-1},\label{eq:fishercov}
\end{equation}
the usual Fisher result.

Now let us imagine splitting our data $\hat{X}$ into two pieces, $\hat{X}_1$ and $\hat{X}_2$, and performing
an alternative parameter analysis using $\hat{X}_1$ alone.    The probability distribution we use for $X_1$, described
by $S_1$, must satisfy
\begin{equation}
e^{-S_1} dX_1 = dX_1 \int dX_2 e^{-S} \label{eq:actint}
\end{equation}
if the two analyses are to be consistent.  Hence overlines for quantities involving $X_1$ alone can equivalently be thought of as 
referring to averages over realisations of $X_1$ alone or
over the full data $X$.  Corresponding to Eq.\ (\ref{eq:dpee}) we have
\begin{equation}
\delta P_1=-\overline{S''_1}^{-1} S'_1, \label{eq:dpee1}
\end{equation}
and corresponding to Eq.\ (\ref{eq:fishercov}) we have
\begin{equation}
\overline{\delta P_1 \delta P_1^T}= \overline{S_1''}^{-1}.
\end{equation}

Now we are in a position to investigate the distribution of parameter \emph{differences}, $\delta P_1-\delta P$, over 
the full ensemble.  With each term averaging to zero, the parameter differences also average to zero.  For the 
covariance, we have
\begin{align}
\overline{\left(\delta P_1-\delta P\right)\left(\delta P_1-\delta P\right)^T}
&=\overline{S''_1}^{-1} \overline{S'_1 {S'_1}^T}  
  \overline{S''_1}^{-1}\nonumber \\ 
&\mbox{}-\overline{S''}^{-1}  \overline{S' {S'_1}^T}   \overline{S''_1}^{-1}\nonumber \\
&\mbox{}-\overline{S''_1}^{-1}   \overline{S'_1 S'^T}   \overline{S''}^{-1}\nonumber \\
&\mbox{}+\overline{S''}^{-1}   \overline{S' S'^T}
  \overline{S''}^{-1} , \label{eq:covexp}
\end{align}
and we see we need the average of the ``mixed'' quantity $S'_1 S'^T$.  This can be obtained as follows:
\begin{align}
\overline{S_{1,i} S_{,j}} &= \int dX S_{1,i} S_{,j} e^{-S} \nonumber \\
&= -\int\ dX S_{1,i}\left(e^{-S} \right)_{,j} \nonumber \\
&= -\int dX\left(S_{1,i}e^{-S} \right)_{,j}+\int dX S_{1,ij} e^{-S}
  \nonumber \\
&= -\int dX_1\left(S_{1,i} \int dX_2 e^{-S} \right)_{,j}+\int dX
  S_{1,ij} e^{-S} \nonumber \\
&= -\left( \int dX_1 S_{1,i} e^{-S_1} \right)_{,j} +\int dX S_{1,ij}
  e^{-S} \nonumber \\
&= 1_{,ij}+\overline{S_{1,ij}} \nonumber \\
&= \overline{S_{1,ij}} ,
\end{align}   
where we have used the fact that by its definition $S_1$ must be independent of $X_2$.  Hence we find
\begin{equation}
\overline{\delta P_1 \delta P^T}= \overline{S''}^{-1},
\labeq{crosscov}
\end{equation}
the same as for $\overline{\delta P \delta P^T}$ itself.
Substituting into Eq.\ (\ref{eq:covexp})
gives us the elegant final result
\begin{equation}
\overline{\left(\delta P_1-\delta P\right)\left(\delta P_1-\delta P\right)^T}=
\overline{S_1''}^{-1} - \overline{S''}^{-1}, \label{eq:covresult}
\end{equation}
i.e.,\ \textit{the covariance of the parameter differences between the
  partial and full analyses is simply the difference of the covariances}.

\section{Interpreting Differences in Multiple Parameters}
\label{sec:multishift}

For multiple parameters, one can form a ``$\chi^2$'' for the differences in the parameters between the two analyses.
This allows one to treat all parameters fairly, neither focussing on one outlier in particular nor neglecting
degeneracies when judging how unlikely multiple shifts are.  If we
write the parameter shifts as $\Delta=\delta P_1 - \delta P$, in the Gaussian approximation, 
\begin{multline}
p(\Delta) d\Delta = \\
\frac{d\Delta}{\left| 2 \pi \(\overline{S_1''}^{-1} - \overline{S''}^{-1} \)\right|^{1/2}}
\exp\left[ - \frac{1}{2} \Delta\tr \(\overline{S_1''}^{-1} -
  \overline{S''}^{-1} \)^{-1} \Delta \right].
\labeq{deltadist}
\end{multline}
If the effective $\chi^2$ (minus twice the exponent) were large then
one might begin to worry about the fidelity of some aspect of the data taken as a whole 
or indeed about the applicability of the model to all of the data.
The former would point to systematic effects, the
latter to new physics.

\section{Example}
\labsec{example}

Let us apply the formalism to an example drawn from cosmic microwave
background (CMB) analysis in cosmology.  Relevant introduction,
motivation and definitions of the model parameters may be found in the Planck ``cosmological parameters'' series of papers 
\citep{2014A&A...571A..16P,2016A&A...594A..13P,2018arXiv180706209P}.

\begin{figure*}
\centering
\includegraphics[width=15cm]{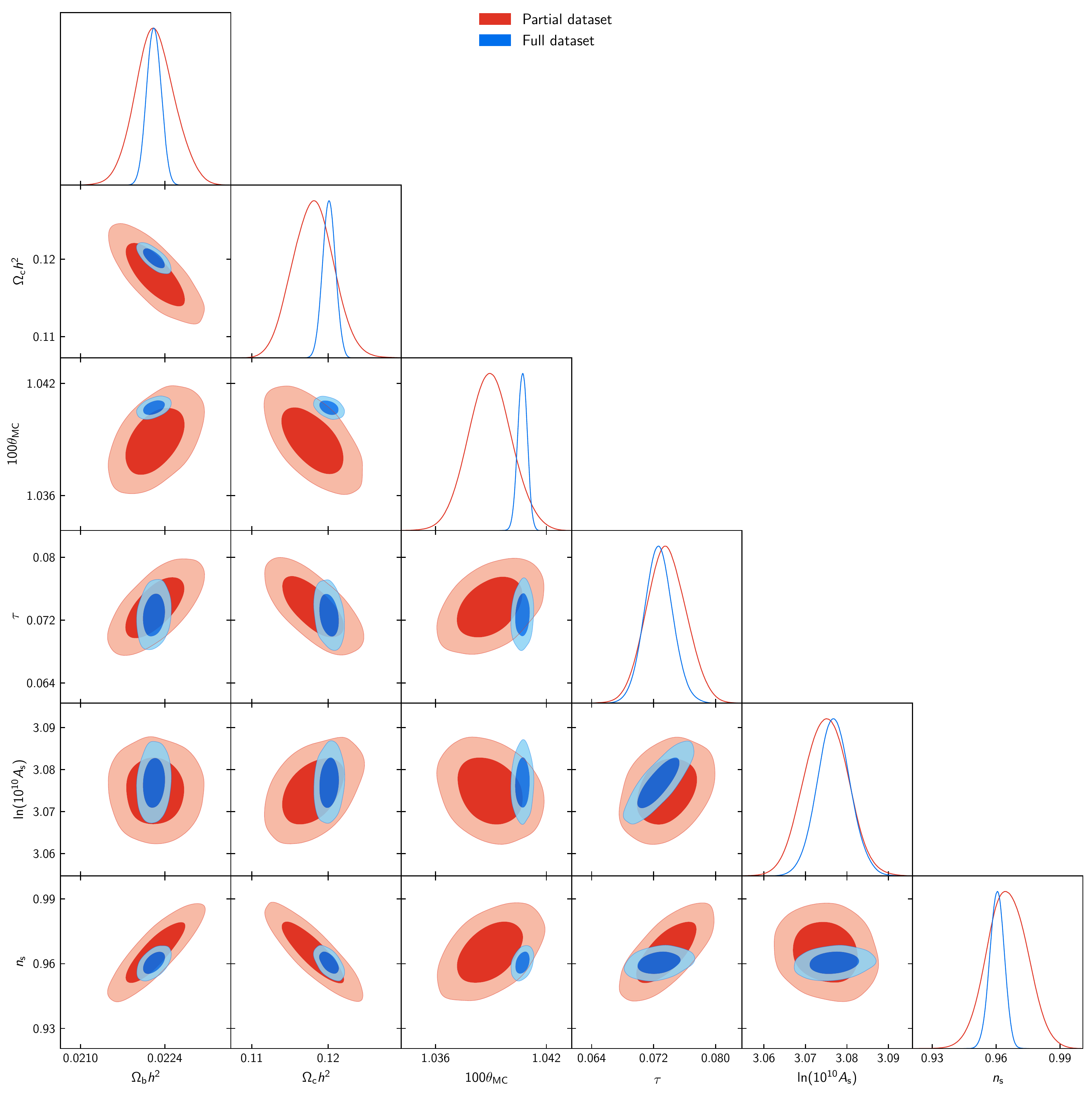}
\caption{
\labfig{fptri} 
Two-dimensional marginalised posterior distributions for a full (blue; smaller contours) and a
partial (red; larger contours) analysis of a simulated CMB dataset.
}  
\end{figure*}

Here one takes maps of the microwave sky and compares them to predictions from a parameterized model, which we take
here to be a standard six-parameter $\Lambda$CDM cosmology.  The model does not predict the actual pattern of fluctuations
of the CMB, only their statistical properties.  The primordial
fluctuations in the early Universe are assumed to be
Gaussianly-distributed, with a 3D power spectrum that changes smoothly
with scale, giving an intrinsic ``cosmic variance'' to observations.
In addition to temperature, ``T'',  or intensity fluctuations, the
linear polarization of the light
also varies across the sky.  In the simplest $\Lambda$CDM models that we shall consider here, this polarization can be described
with the help of an additional scalar field, ``E''.  From the T and E fields, there are three angular power spectra we can construct, the two auto-spectra
$\hat{C}^\mathrm{TT}_\ell$ and $\hat{C}^\mathrm{EE}_\ell$ and the cross-spectrum $\hat{C}^\mathrm{TE}_\ell$ ($\ell$ denoting the angular multipole number).  Neglecting Galactic and extra-galactic contamination, and assuming the full sky is observed with no instrument noise, the negative of the log-likelihood of the observed spectra is deducible to be
\begin{equation}
-\ln L = \(l+1/2\)\[\text{tr} \left(\mbC_\ell^{-1} \mbhatC_\ell\right) + \ln\frac{\left| \mbC_\ell \right|}{\left| \mbhatC_\ell \right|} -1 \], \labeq{scorrexact}
\end{equation}
where 
\begin{equation}
\mbhatC_\ell =
\begin{pmatrix}
\hat{C}^\mathrm{TT}_\ell & \hat{C}^\mathrm{TE}_\ell \\
\hat{C}^\mathrm{TE}_\ell & \hat{C}^\mathrm{EE}_\ell 
\end{pmatrix}
\end{equation}
and $\mbC_\ell$ is defined similarly but with the theory spectra.

Now it might be that an experiment is such that the polarization is observable only on larger angular scales (lower $\ell$), whereas temperature measurements are possible
down to finer angular scales (higher $\ell$).  So we shall here
investigate the shifts anticipated between a ``full'' measurement with
TT, TE and EE spectra for angular multipoles $2 \leq \ell \leq 800$
and a ``partial" one retaining the range  $2 \leq \ell \leq 800$ for
TT but only $2 \leq \ell \leq 29$ for TE and EE.  (While somewhat
arbitrary, these choices for the limits have been chosen to correspond
to, e.g.,  the investigation of dependence of the parameters on
multipole cuts in \citealt{2019arXiv191000483E}.) This makes for a good test of our formalism since theory predicts the temperature and polarization signals to 
be correlated with each other. 

We generate a fiducial power spectrum from a ``best-fit'' model of the
Planck 2015 analysis~\citep{2016A&A...594A..13P} using the {\tt CAMB} \citep{2000ApJ...538..473L}
software.  Next, we generate a realization of the TT, TE and EE
spectra from this model by first drawing Gaussian realizations of the
T and E multipoles and forming their auto- and cross-power spectra.
Finally, we
perform Markov-chain Monte-Carlo (MCMC) analyses on both the full set of spectra and the partial
set and find best-fit models in both cases using the {\tt CosmoMC}
\citep{2002PhRvD..66j3511L} software. Parameter means, standard deviations and
best-fits are listed in Table \ref{tab:costab} and 2D marginalised
posterior distributions are illustrated in Fig.\ \reffig{fptri}.

\begin{table*}
\caption{
\label{tab:costab}
Cosmological parameter constraints from a full and partial analysis of
a simulated CMB data set (the posterior distributions are shown in Fig.~\reffig{fptri}).
}\centering
\begin{tabular}{|l|c|c|c|c|c|c|}

\toprule
						&\multicolumn{3}{|c|}{Full analysis}&\multicolumn{3}{|c|}{Partial analysis}\\
\midrule
						&best fit		& mean		&std.\ dev.	 	& best fit 		& mean		&std.\ dev.  \\
\midrule
$\Omega_b h^2$ 			&0.02221		&0.02222		&0.00012		&0.02219		&0.02224		&0.00032	\\
$\Omega_c h^2$ 			&0.12013		&0.12011		&0.00083		&0.11844		&0.11804		&0.00267	\\
$100 \Theta_\mathrm{MC}$	&1.04070		&1.04071		&0.00025		&1.03886		&1.03895		&0.00118	\\
$\tau$					&0.07255		&0.07265		&0.00183		&0.07337		&0.07365		&0.00254	\\
$\ln(10^{10} A_s)$ 			&3.07673		&3.07689		&0.00399		&3.07529		&3.07498		&0.00522	\\
$n_s$					&0.96030		&0.96040		&0.00335		&0.96384		&0.96522		&0.00937	\\
\bottomrule
\end{tabular}

\end{table*}

The {\tt CosmoMC} software also provides estimates of the covariance
matrices for the posterior distributions derived from the MCMC chains.  Using
these in Eq.\ \refeq{covresult}, we can derive standard deviations for the \emph{shifts} in parameters between the full and
partial analyses.  We list in Table \ref{tab:cosshift} these standard
deviations, along with the measured shifts in both the best-fits and the
means in terms of these standard deviations.

\begin{table}
\caption{
\label{tab:cosshift}
Expected standard deviations of the shifts in best-fitting parameters
between the partial and full analyses of the
simulated CMB dataset reported in Table~\ref{tab:costab}
(first column;
computed from Eq.~\ref{eq:covresult}) compared to the measured shifts in best-fitting parameters
(in units of the expected standard deviation of the shift; second
column) and the measured shifts in posterior means (third column).
}
\centering
\begin{tabular}{|l|c|c|c|}
\toprule
						&std.\ dev.\	& $\Delta$(best fit) in &$\Delta$(mean) in	 \\
						&of shift 		& shift std.\ dev.'s	&in shift std.\ dev.'s  	\\
\midrule
$\Omega_b h^2$ 			&0.00030		&$-0.10$	&$-$0.06		\\
$\Omega_c h^2$ 			&0.00254		&$-0.67$	&$-$0.82		\\
$100 \Theta_\mathrm{MC}$	&0.00115		&$-1.59$	&$-$1.53		\\
$\tau$					&0.00175		&$+0.47$	&$+$0.57		\\
$\ln(10^{10} A_s)$ 			&0.00337		&$-0.43$	&$-$0.56		\\
$n_s$					&0.00875		&$+0.40$	&$+$0.55		\\
\bottomrule
\end{tabular}

\end{table}

Considering more than one parameter at a time, one might perform a singular-value-decomposition of Eq.\ \refeq{covresult} to identify the ``most likely'' shifts one should expect to see.  One can also use the entire covariance of Eq.\ (\ref{eq:covresult}) as discussed in Sec.\ \refsec{multishift} to compute
an overall ``goodness-of-fit'' for the shifts in all of the parameters.  We obtain a $\chi^2$ of 9.84 for the shifts in the means (11.3 for shifts in best-fits) for our six degrees of
freedom, a value greater than which would be expected about $13\%$
($8\%$) of the time under the distribution in Eq.~\refeq{deltadist}.  

In computing the standard deviations of parameter shifts shown in Table \ref{tab:cosshift} and associated $\chi^2$ values, we have used the covariances estimated from the MCMC chains, in a similar manner to what one would need to do in a real problem.  In our simulation here, however, we know what the underlying model is and so can calculate $\overline{S_{,ij}}$ and $\overline{S_{1,ij}}$ analytically in terms of derivatives of the fiducial spectra with respect to the model parameters (these derivatives being evaluated numerically).  Using these matrices
the standard deviations of the shifts change very little, but the $\chi^2$ of the shift in the means changes to 8.93 and the  $\chi^2$ of the shift in the best-fits changes to 9.93.

\begin{figure}
\centering
\subfloat{\includegraphics[width=\columnwidth]{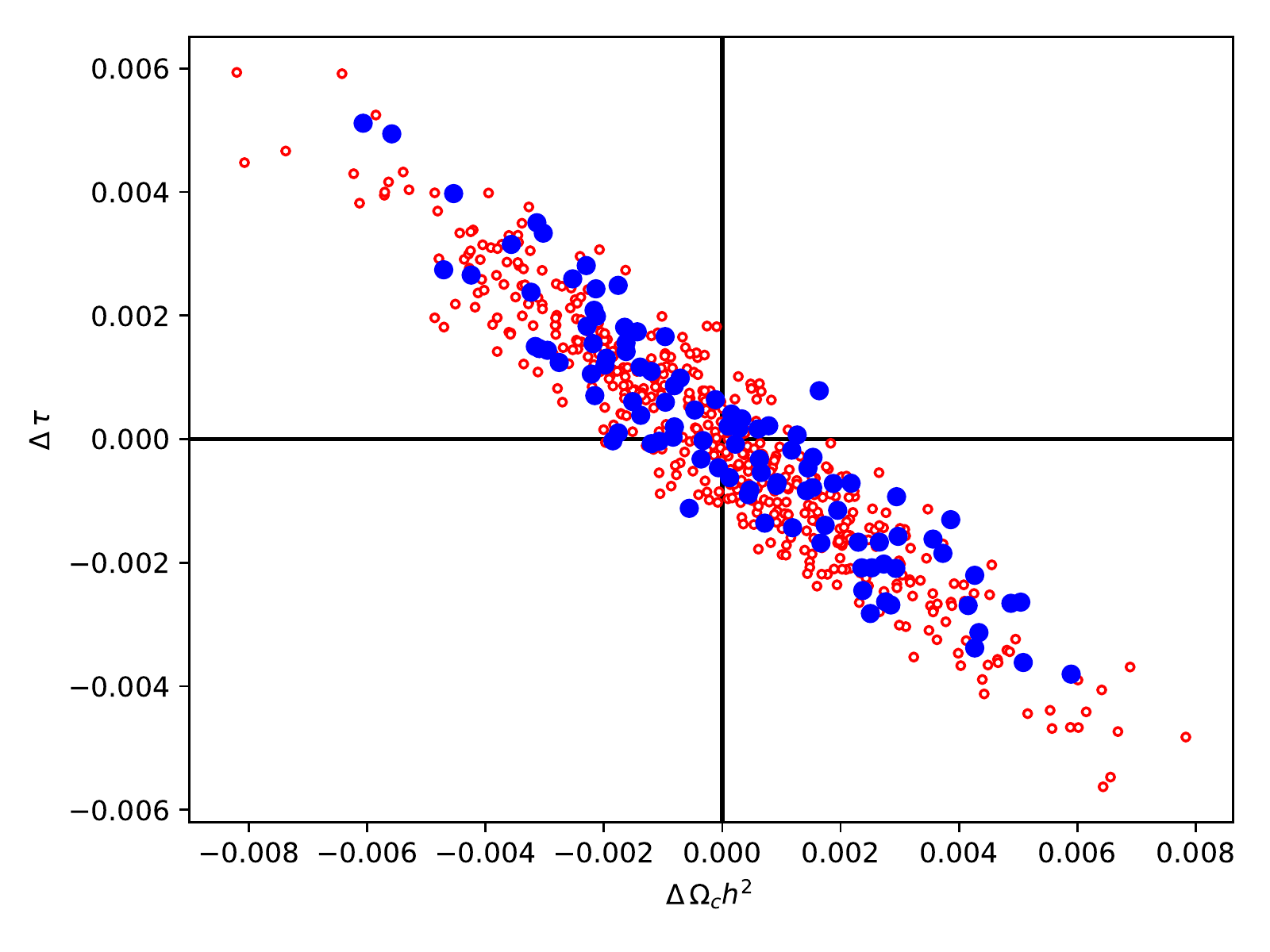}} \\
\subfloat{\includegraphics[width=\columnwidth]{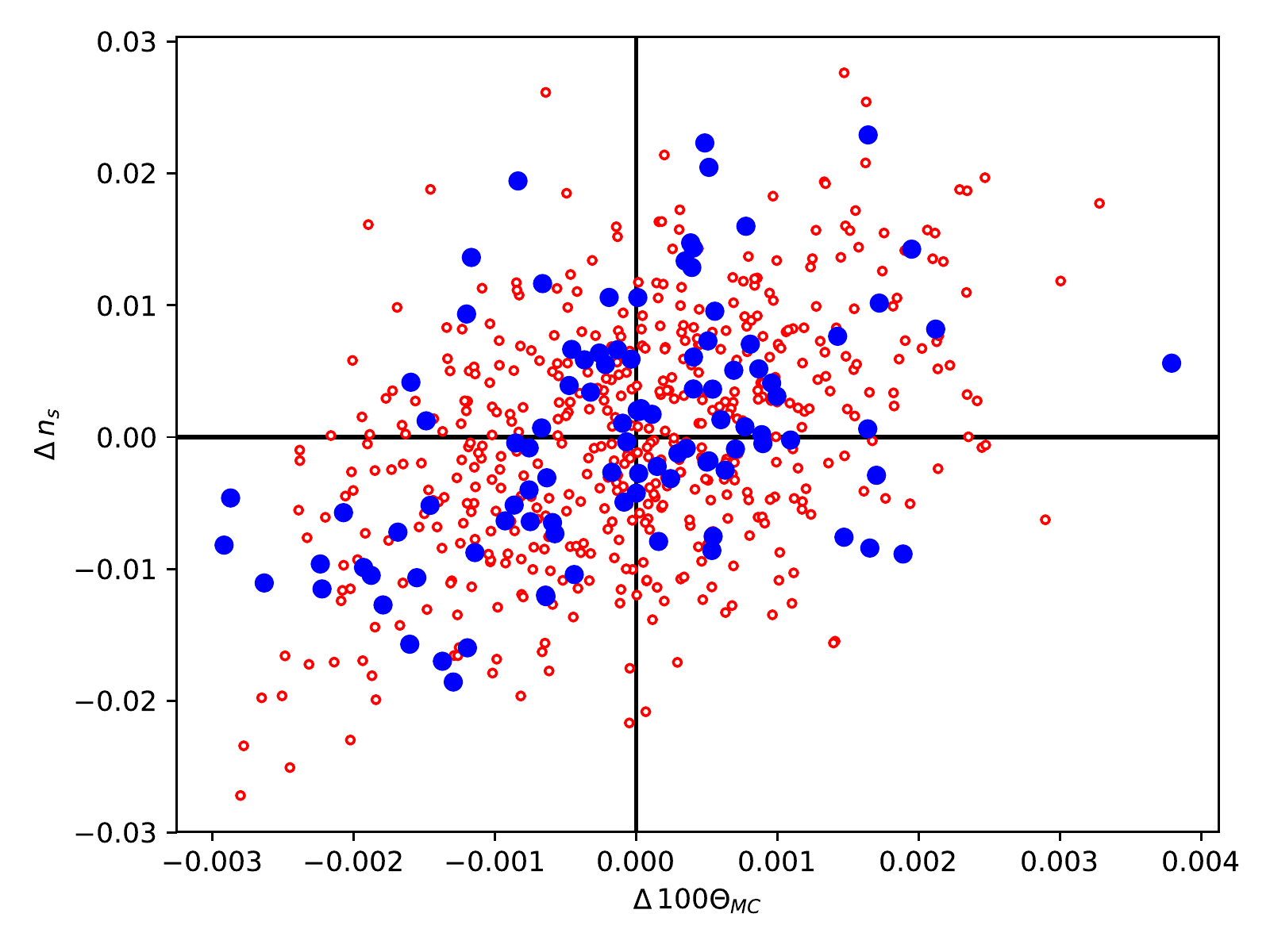}} \\
\subfloat{\includegraphics[width=\columnwidth]{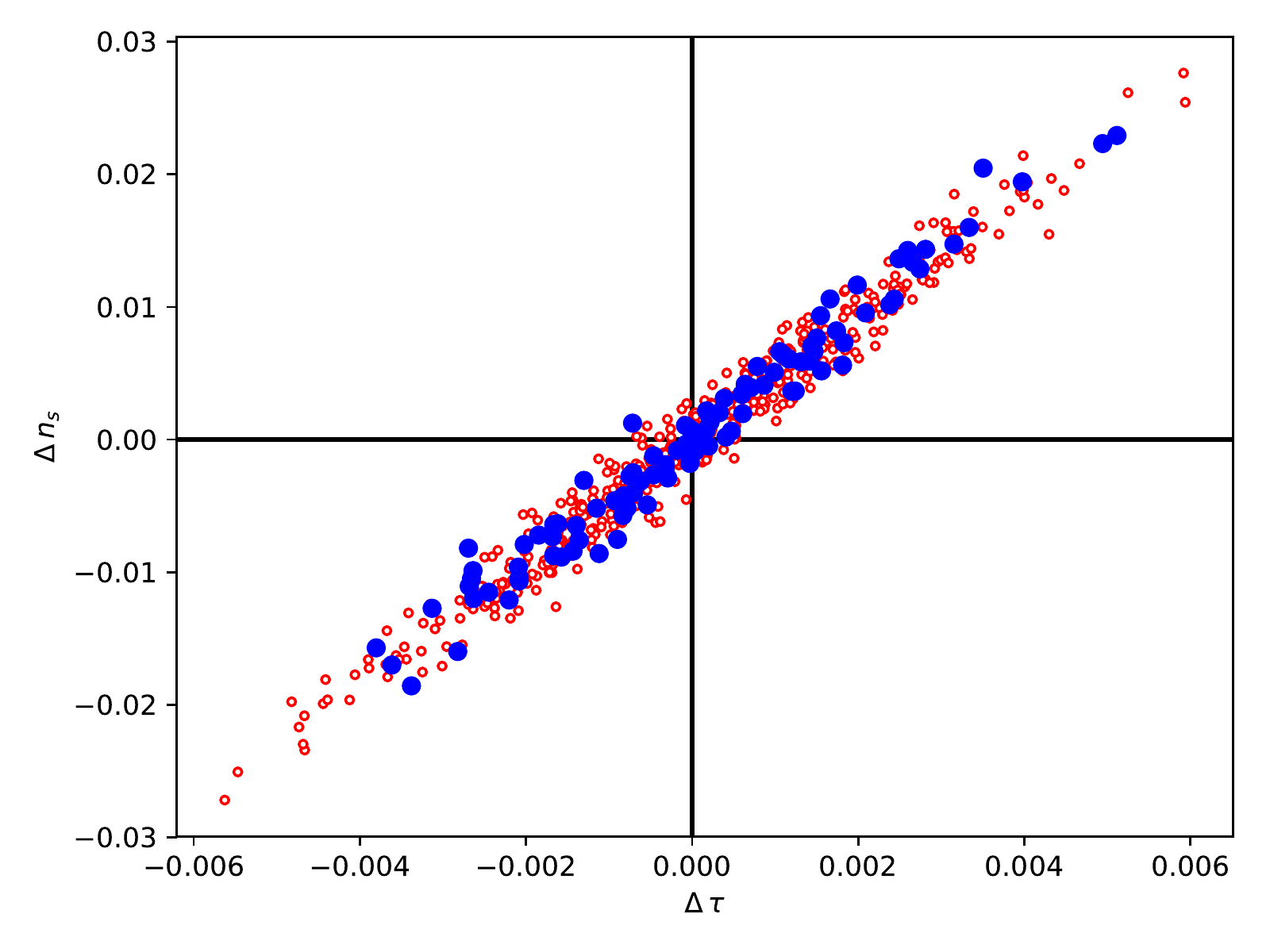}}  
\caption{
\labfig{psillus} 
Representative plots showing shifts in pairs of parameters between the
partial and full analyses of 100 simulated CMB datasets (large blue filled circles),
compared to those expected from Eq.\ (\ref{eq:covresult}) (illustrated via 500 Gaussian realizations displayed with small red open circles).    
}  
\end{figure}

We go on to generate 100 further realizations of the power spectra, and find their best-fitting parameter values under the full and partial treatments.  We compare the shifts between the analyses to predictions from Eq.\ (\ref{eq:covresult}) in Fig.\ \reffig{psillus}.  We plot
a histogram of the $\chi^2$s of the shifts, using the analytic covariances, 
in Fig.\ \reffig{chisqfig_minierror}.  In order to obtain the good agreement with expectation shown, note that we had to add terms to the covariance to account for the 0.05-sigma tolerances in the minimization procedure used.  We did this by adding $0.05^2$ times the diagonals of each of the full and partial covariances back to the covariance of the difference; without these terms a small number of the realizations appeared to have very unlikely shifts.  Thus it would seem prudent to consider the potential effect of including similar terms if initially faced with a high $\chi^2$ from parameter shifts estimated from either best-fits or means in some analysis.

\begin{figure}
\centering
\includegraphics[width=\columnwidth]{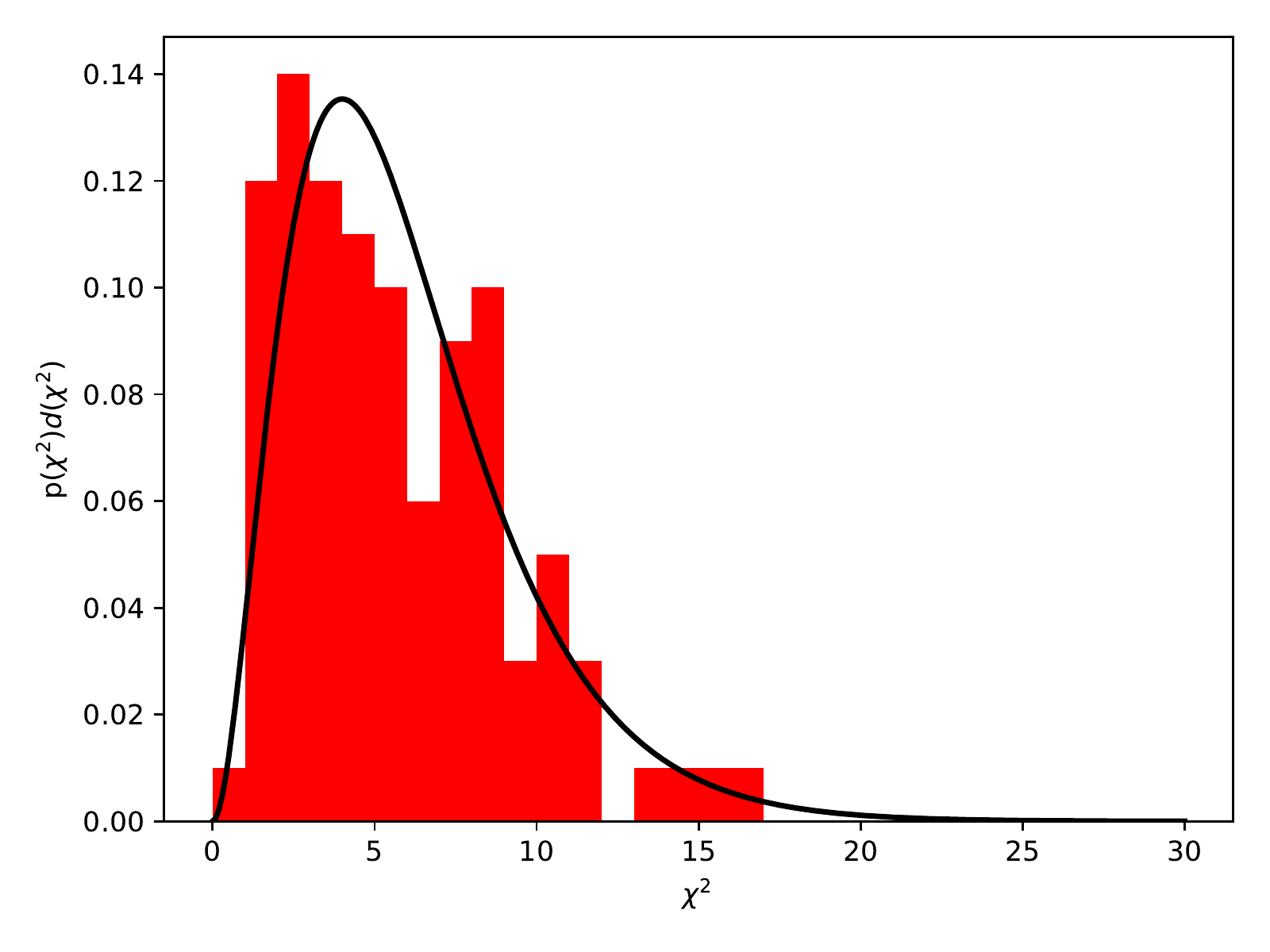}
\caption{
\labfig{chisqfig_minierror} 
Normalized histogram showing the effective $\chi^2$ from Eq.\ \refeq{deltadist} evaluated for the difference between the partial and full
analyses of 100 simulated CMB datasets, using analytic covariances computed around the fiducial model (including terms accounting for tolerances in the minimization procedure), compared to a $\chi^2$ distribution for six degrees of freedom.
}  
\end{figure}

\section{Comments and Extensions}

Our result Eq.\ \refeq{covresult} should have a wide applicability,
enabling one to compare analyses with differing combinations of
datasets, e.g., to illuminate
tensions between late-time measurements of the expansion rate of the Universe (see, e.g., \citealt{2018ApJ...855..136R}) and inferences from the $\Lambda$CDM framework with Planck (see, e.g., the discussion in \citealt{2018arXiv180706209P}),  as well as comparing subsets of data taken with the same experiment.

We can understand generic features of parameter shifts from the form of Eq.\ \refeq{covresult}.  For example, if large parameter degeneracies that exist using the partial data set are broken using the full data set, then one would expect parameter differences to lie along those parameter degeneracies also.

Note that in order to obtain the simple result of Eq.\ (\ref{eq:covresult}), we had to look at differences between one data 
combination and another ``nested'' within it.  No such simple result exists in general for parameter
differences between, say,  two non-nested datasets $X_1$ and $X_2$.  If the data sets happen to be independent, then our method does yield 
\begin{equation}
\overline{\left(\delta P_1-\delta P_2\right)\left(\delta P_1-\delta P_2\right)^T}
=
\overline{S_1''}^{-1} + \overline{S_2''}^{-1},
\end{equation}
 with uncertainties now adding in quadrature as expected.

By applying Eq.\ (\ref{eq:covresult}) to multiple nested subsets, one can build up a ``grand'' covariance matrix
for expectations of the parameter differences between all the analyses.  Let parameters $P_2$ come from an analysis involving
a subset of the data that yielded parameters $P_1$, itself from an analysis involving a subset of the data that yielded 
parameters $P$.  
By considering appropriate conditional distributions 
associated with this covariance matrix, certain properties of the parameters obtained may be understood.  For example,
using Eqs.\ \refeq{fishercov} and \refeq{crosscov} we can compute the joint covariance of the differences 
$P_1-P$ and $P_2-P$ to compare their behaviour to expectation.  

Knowing the joint distribution for the fluctuations in parameters from the truth (i.e.\ the $\delta P$, $\delta P_1$, \ldots, themselves), we can 
understand certain properties of the  behaviour of the parameters by considering the associated conditional distributions.  
For example, given $\delta P$, one's best estimate for  $\delta P_1$ is just  $\delta P$, whereas given $\delta P_1$, one 
should expect $\delta P$ to be $ C C_1^{-1} \delta P_1$ (with the $C$s
denoting respective covariance matrices).  For a single parameter this
reduces to $\sigma^2 / \sigma_1^2$ times $\delta P_1$; as more data is
added and uncertainties reduce, parameters are most likely to move
monotonically closer to the truth.
Given $P$ and $P_2$, one can show that $P_1$ should peak around an intermediate value between the two.  We can also gain
some intuition about how the $\chi^2$ of fits should behave.  Focusing
on a single parameter for simplicity, given some $\delta P_1$ with a
$\chi^2_1=\delta P_1^2 / \sigma_1^2 $, we should conditionally expect
the fuller analysis to have a $\chi^2$ of $\sigma^2 \chi^2_1 /
\sigma_1^2 + (\sigma_1^2-\sigma^2) / \sigma_1^2$; if additional data
is added that significantly reduces uncertainties, the significance of
an outlier should decrease.

Finally, one can generalise the argument of Sec.\ \refsec{thearg} to include Gaussian priors on the parameters. However, the result is not
as easily expressed in terms of the covariance matrices involved as it was in Eq.\ (\ref{eq:covresult}). 

\section{Conclusions}

In this note we have presented arguments aimed to help the understanding of relations 
between inferences using full and partial amounts of data and, in the appendix, between
inferences using standard and extended models. 

Our work provides some insight into the sorts of posterior 
variations one might expect when comparing related Bayesian parameter estimations.  A ``rule of 
thumb'' for a parameter that
is well-constrained by the data (so that any effect of priors may be neglected) is as follows: when more data is
added, a significant change in the width of the  
posterior distribution will be accompanied by a difference in the most
likely parameter value that can be a significant fraction of the
larger width. 
If the posteriors
have similar widths then there should be little shift in the peak position.

\section*{Acknowledgements}

We thank George Efstathiou, Antony Lewis and the Planck Parameters and Likelihood teams
for helpful comments and discussions over the development of this work. 

Our MCMC analyses were performed using the Cambridge Service for
  Data Driven Discovery (CSD3) operated by the University of Cambridge
  Research Computing Service (\url{http://www.csd3.cam.ac.uk/}),
  provided by Dell EMC and Intel using Tier-2 funding from the
  Engineering and Physical Sciences Research Council, and DiRAC
  funding from the Science and Technology Facilities Council
  (\url{www.dirac.ac.uk}).

SG and AC acknowledge support from the UK Science and Technology
  Facilities Council (grant numbers ST/N000927/1 and ST/S000623/1).

\appendix
\section{Wilks' Theorem}

Imagine we fit a model with $n_a$ parameters to our data, and then
fit an extended model with an additional $n_b$ parameters. In the
case where the first model is the correct one, i.e., the additional
parameters are not required, by how much should we expect the extended
model to improve the goodness of fit by chance?
For a cosmological example, one might allow the amplitude of the 3D power
spectrum of primordial gravitational waves to vary in the extended
model when they are actually negligible.
Wilks' Theorem \citep{wilks}
leads one to expect the improvement in the log-likelihood for the wider analysis over the more specific one to be $\chi^2$-distributed
with a number of degrees of freedom equal to the number of additional parameters the wider analysis has compared to the
more specific one (i.e., $n_b$).

Equations~\refeq{actexp} and \refeq{dpee} can be used to rederive this result. Here we
need to keep track of the constant to evaluate $S$ at the best-fit model:
\begin{align}
S_\mathrm{bf}&\approx S_0-  \frac{1}{2} {S'}\tr
               \overline{S''}^{-1} S' \nonumber \\
&= S_0 - \frac{1}{2} \delta P\tr \overline{S''}  \delta P ,
\end{align}
where we understand the parameter shifts $\delta P$ in the second line here to be the difference between the best-fit model
of the class considered and the underlying one.

We split our $\delta P$ into two parts,  $\delta P_a$ and  $\delta P_b$,
corresponding to the usual parameters and the additional parameters, respectively, of lengths $n_a$ and $n_b$.  For the 
restricted analysis, we have
\begin{equation}
 \delta P^\mathrm{R}_a=-\overline{S''_{aa}}^{-1} S'_a ,
\end{equation}
 where we have partitioned $S'$ and $S''$ as
 \begin{align}
 S'&=
 \begin{pmatrix}
 S'_a \\
 S'_b
 \end{pmatrix} , \\
 S''&=
 \begin{pmatrix}
 S''_{aa} & S''_{ab} \\
 S''_{ba} & S''_{bb}
 \end{pmatrix}.\labeq{sppdecomp}
 \end{align}
 Introducing an $n$-by-$n_a$ projection matrix $M$ (with $n=n_a+n_b$,
 corresponding to the total number of parameters varied in the wider analysis),
 \begin{equation}
 M=
 \begin{pmatrix}
 I \\
 0 
 \end{pmatrix}
 \end{equation}
we can express $S_\mathrm{bf}-S_0$ for the usual model as
 \begin{align}
 S_{a\mathrm{bf}}-S_0
 &=-\frac{1}{2}(\delta P^\mathrm{R}_a)\tr \overline{S''_{aa}}
   \delta P^\mathrm{R}_a \nonumber \\
 &=-\frac{1}{2} {S'_a}\tr \overline{S''_{aa}}^{-1} S'_a
   \nonumber \\
&=-\frac{1}{2} {S'}\tr M \overline{S''_{aa}}^{-1} M\tr S' .
\end{align}
Subtracting this from the same quantity evaluated for the extended model yields
\begin{align}
S_\mathrm{bf}-S_{a\mathrm{bf}}&= -\frac{1}{2} {S'}\tr  
\( \overline{S''}^{-1} - M \overline{S''_{aa}}^{-1} M\tr \)
S' \nonumber \\
&= -\frac{1}{2} {\delta P}\tr  
\( \overline{S''} -  \overline{S''} M \overline{S''_{aa}}^{-1} M\tr \overline{S''}  \)
\delta P
\end{align}
($\delta P$ here being the shift from the underlying model to the extended best-fit model).
Using Eq.~\refeq{sppdecomp} the bracketed term becomes
\begin{equation}
\begin{pmatrix}
 0 & 0 \\
 0 & \overline{S''_{bb}} -\overline{S''_{ba}} \, \overline{S''_{aa}}^{-1} \overline{S''_{ab}}
 \end{pmatrix}
=
\begin{pmatrix}
 0 & 0 \\
 0 & \({\overline{S''}^{-1}}_{bb}\)^{-1}
 \end{pmatrix}.
 \end{equation}
 Hence
 \begin{equation}
S_\mathrm{bf}-S_{a\mathrm{bf}}
= -\frac{1}{2} {\delta P_b}\tr 
\({\overline{S''}^{-1}}_{bb}\)^{-1}
 \delta P_b, \labeq{dchi2}
 \end{equation}
 which we note only depends on the additional parameters $\delta P_b$.  So to understand how this is 
 distributed, we need to know how the $\delta P_b$ are distributed.  
 
From Eq.\ \refeq{fishercov}, we see that the $\delta P$ have
covariance $\overline{S''}^{-1}$, and hence the  $\delta P_b$ have covariance
${\overline{S''}^{-1}}_{bb}$, the same matrix as appears in
  the right-hand side of Eq.~\refeq{dchi2}.
So, to the extent that the parameter
shifts may be approximated as Gaussians about the fiducial model, we
can immediately recognize  
$-2\(S_\mathrm{bf}-S_{a\mathrm{bf}}\)$ to be $\chi^2$-distributed with $n_b$ degrees of freedom as in \citet{wilks}.

We may use some of the above to investigate the distribution of shifts between parameters
between the extended and standard analysis. Let us set 
\begin{equation}
\Delta = \delta P - 
\begin{pmatrix}
\delta P^\mathrm{R}_a \\
0
\end{pmatrix},
\end{equation}
i.e., the difference in parameters between the two analyses, and see how $\Delta$ is distributed.
One finds
\begin{equation}
\Delta=-\(  \overline{S''}^{-1}-M \(M\tr  \overline{S''} M \)^{-1} M\tr  \) S' \labeq{projsp}
\end{equation}
and hence, using Eq.~\refeq{fishercov}, we have
\begin{equation}
\overline{\Delta \Delta\tr } =
\overline{S''}^{-1}-M \(M\tr  \overline{S''} M \)^{-1} M\tr .\labeq{projcov}
\end{equation}
The right-hand side of Eq.~\refeq{projcov} is not full rank, constraining $\Delta$ to lie in a subspace of dimension $n_b$; from Eq.~\refeq{projsp} 
it is evident that $\Delta$ is a projection of $\delta P$. 

\bibliographystyle{mnras}
\bibliography{paramshifts.bib}

\bsp 
\label{lastpage}
\end{document}